\documentclass[aps,prl,reprint,superscriptaddress,nofootinbib,floatfix]{revtex4-2}

\usepackage[T1]{fontenc}
\usepackage[utf8]{inputenc}
\usepackage{amsmath,amssymb}
\usepackage{graphicx}
\usepackage{xcolor}
\usepackage{hyperref}
\usepackage{booktabs}

\hypersetup{
  colorlinks=true,
  linkcolor=blue,
  citecolor=blue,
  urlcolor=blue
}

\begin{document}

\title{Searching for Continuous Gravitational Waves with Gaia Astrometric Accelerations}

\author{Song Li}
\affiliation{Shanghai Astronomical Observatory, Shanghai, 200030, China}
\email{leesong@shao.ac.cn}

\author{Jinglong~Liu}
\affiliation{State Key Laboratory of Dark Matter Physics, Tsung-Dao Lee Institute \& School of Physics and Astronomy, Shanghai Jiao Tong University, Shanghai 201210, China}
\affiliation{School of Physics and Astronomy, Shanghai Jiao Tong University, 800 Dongchuan Road, Shanghai 200240, China}
\email{jinglong\_liu@sjtu.edu.cn}

\author{Hiromichi Tagawa}
\affiliation{Shanghai Astronomical Observatory, Shanghai, 200030, China}
\email{htagawa@shao.ac.cn}

\author{Shu-Cheng Yang}
\affiliation{Shanghai Astronomical Observatory, Shanghai, 200030, China}
\email{ysc@shao.ac.cn}

\author{Fabo Feng}
\affiliation{State Key Laboratory of Dark Matter Physics, Tsung-Dao Lee Institute \& School of Physics and Astronomy, Shanghai Jiao Tong University, Shanghai 201210, China}
\email{ffeng@sjtu.edu.cn}

\date{\today}

\begin{abstract}
Gravitational waves (GWs) induce coherent, time-dependent deflections in the apparent positions of distant sources. Previous astrometric searches have used quasars to constrain the resulting proper-motion patterns. In this Letter, we present the first search for continuous GWs from individual supermassive black hole binaries using angular-acceleration estimates derived from stellar proper-motion differences between Gaia DR2 and DR3. For GW periods long compared with Gaia observing windows, the angular-acceleration response increases more rapidly toward shorter periods than the proper-motion response. We derive 95\% credible upper limits on the strain amplitude under several astrometric-systematics treatments. The model including magnitude- and color-dependent frame rotation and a global glide yields the tightest observed limit among those tested, approximately $4\times10^{-11}$ near $f_{\mathrm{GW}}=10\,\mathrm{nHz}$. These results highlight the importance of modeling differential astrometric systematics in future astrometric GW searches.
\end{abstract}

\maketitle

\textit{\textbf{Introduction---}}
The detection of GW150914 by the Laser Interferometer Gravitational-Wave Observatory (LIGO) marked the first observation of gravitational waves (GWs) from a binary black hole merger \cite{GW150914}. Recent detections reported in version 5.0 of the Gravitational-Wave Transient Catalog (GWTC-5.0) by the LIGO–Virgo–KAGRA (LVK) Collaboration have reported 390 candidates with an astrophysical-origin probability $p_{\rm astro}\geq0.5$ in the frequency band \cite{LIGO_7}. Compared with ground-based observations, lower-frequency GW measurements probe more massive black hole binaries at earlier stages of their evolution. The Laser Interferometer Space Antenna (LISA) will explore the millihertz band \cite{LISA2024}, while pulsar timing arrays (PTAs) search for nanohertz GWs \cite{Zic2023PPTADR3, Agazie2023Timing, Antoniadis2023EPTAData, Rana2025InPTADR2, Chen2025CPTANoise, Miles2025MPTAData}. PTAs have found evidence for a stochastic GW background with Hellings--Downs spatial correlations, consistent with a population of supermassive black hole binaries (SMBHBs) \cite{NANOGrav2023, Agazie2023GWB, Reardon2023GWB, Antoniadis2023EPTAGWB, Xu2023CPTAGWB, Miles2025MPTAGWB}. These observations also inform models of SMBHB populations and evolution \cite{Jaffe2003, Wyithe2003, Enoki2004, Sesana2008}. Complementary searches for continuous GWs from individual binaries have found no compelling detections but have set frequency-dependent strain limits \cite{Babak2016CW, Agazie2023CW, Falxa2023CW, Zhao2025CW}. A recent redshift-aware PTA analysis has further constrained high-redshift sources, reaching source-frame chirp mass upper limits of a few $10^9\, M_{\odot}$ in its most sensitive frequency band \cite{Chen2026}.

Astrometry offers a complementary way to search for individual continuous-wave sources in the nanohertz band. A passing GW perturbs the apparent directions of distant sources, producing a coherent pattern of angular deflections across the sky. The time dependence of these deflections induces correlated proper motions and angular accelerations, allowing an ensemble of distant sources to act as a GW detector \cite{Braginsky1990, Pyne1996, BookFlanagan2011, Moore2017, Klioner2018}. Quasars are particularly useful because their cosmological distances strongly suppress proper motions arising from transverse peculiar velocities. Their broad sky coverage also enables the GW-induced spatial pattern to be distinguished from reference-frame rotation and other large-scale astrometric systematics.

Gaia’s all-sky astrometry has enabled observational constraints on GW-induced astrometric correlations. Using Gaia Data Release 3 (DR3) quasar proper motions, Jaraba et al. constrained the stochastic GW background through joint dipole and quadrupole fits \cite{Jaraba2023}. Darling subsequently obtained tighter limits using a larger quasar sample and an astrometric Hellings--Downs correlation analysis \cite{Darling2025}. These studies constrain the frequency-integrated power of a stochastic background. Searches for individual continuous-wave sources instead exploit the coherent angular pattern and temporal evolution associated with a specific GW propagation direction \cite{Moore2017}. Filipello \cite{Filipello_26} showed that a joint analysis of quasars and several high-precision pulsars can improve upper limits on continuous gravitational waves (CGWs). Gaia also provides astrometry for a vastly larger stellar population. The greater number of stars and their higher astrometric precision will enable stronger constraints on individual GW sources. Also, if we can make use of the two Gaia data releases (DR2 and DR3), it might give a better constraint on GWs. 

In this Letter, we present the first search for continuous GWs using angular-acceleration estimates derived from the proper-motion differences of Gaia DR2 and DR3 stars. This observable adds temporal information to searches based on a single catalog's proper-motion field. When analyzing the data, we consider the effects of frame rotation and glide and find that frame rotation is the most important effect when constraining the GWs.
We find no significant GW evidence and report upper limits on the strain over the considered period range.

\textit{\textbf{Astrometric response to GWs---}}
A passing GW perturbs the light path from a source to the observer, changing the apparent direction of the source on the sky. Braginsky proposed this astrometric effect as a means of detecting GWs \cite{Braginsky1990}; Pyne derived the deflection of a distant source \cite{Pyne1996}, and Book and Flanagan developed the response and its correlated sky pattern \cite{BookFlanagan2011}. For a single continuous-wave source, the resulting time-dependent deflections form a coherent angular pattern across the observed sources. In this work, we use the observer-term response to model that pattern.

In the distant-source approximation, the coherent observer term produces an angular deflection,
\begin{equation}
\delta n_a^{\rm GW}(t)
=
\frac{n_a-q_a}
{2(1-\hat{\boldsymbol q}\cdot\hat{\boldsymbol n})}
h_{bc}(t)n^b n^c
-\frac{1}{2}h_{ab}(t)n^b,
\label{eq:gw-deflection}
\end{equation}
where $\hat{\boldsymbol n}$ is the unperturbed unit direction to the observed source, $\hat{\boldsymbol q}$ points toward the GW source, and $h_{ab}(t)$ is the dimensionless GW metric perturbation evaluated at the observer. The GW propagates along $-\hat{\boldsymbol q}$, and repeated spatial indices are summed. We neglect the distance-dependent source term in the coherent full-sky analysis. When $D f_{\rm GW}/c\gg1$, its phase varies rapidly with source distance and sky position, so contributions from sources at different distances are expected to average out when the measurements are combined. 

For a circular binary whose evolution is negligible over the observing span, we write
\begin{equation}
h_{ab}(t)=A_+ e^+_{ab}(\hat{\boldsymbol q},\psi)\cos\Phi(t)
 +A_\times e^\times_{ab}(\hat{\boldsymbol q},\psi)
 \sin\Phi(t),
\label{eq:gw-metric}
\end{equation}
$\Phi(t)=2\pi f_{\rm GW}(t-t_0)+\phi_0$, $A_+=\frac{A}{2}(1+\cos^2\iota), A_\times=A\cos\iota.$ Here, $A$ sets the strain amplitude, $\iota$ is the binary inclination, $\phi_0$ is the initial phase angle, and $\psi$ is the polarization angle. In the following, we use $f$ as shorthand for the GW frequency $f_{\mathrm{GW}}$.

The observable angular displacement is obtained by projection onto the local right-ascension and declination basis, $\boldsymbol{\xi}_i^{\rm GW}(t)
= \bigl(\hat{\boldsymbol e}_{\alpha,i}\cdot\delta\boldsymbol n_i^{\rm GW}(t),
\hat{\boldsymbol e}_{\delta,i}\cdot\delta\boldsymbol n_i^{\rm GW}(t)\bigr)^{\mathsf T}$, the first component corresponds to $\delta\alpha_i\cos\delta_i$. 

For the monochromatic signal, the instantaneous angular acceleration is
\begin{equation}
 \boldsymbol a_i^{\rm GW}(t)
 =
 \ddot{\boldsymbol\xi}_i^{\rm GW}(t)
 =
 -(2\pi f)^2
 \boldsymbol\xi_i^{\rm GW}(t).
 \label{eq:gw-acceleration}
\end{equation}

\begin{figure}[t]
  \centering
  \includegraphics[width=\columnwidth]{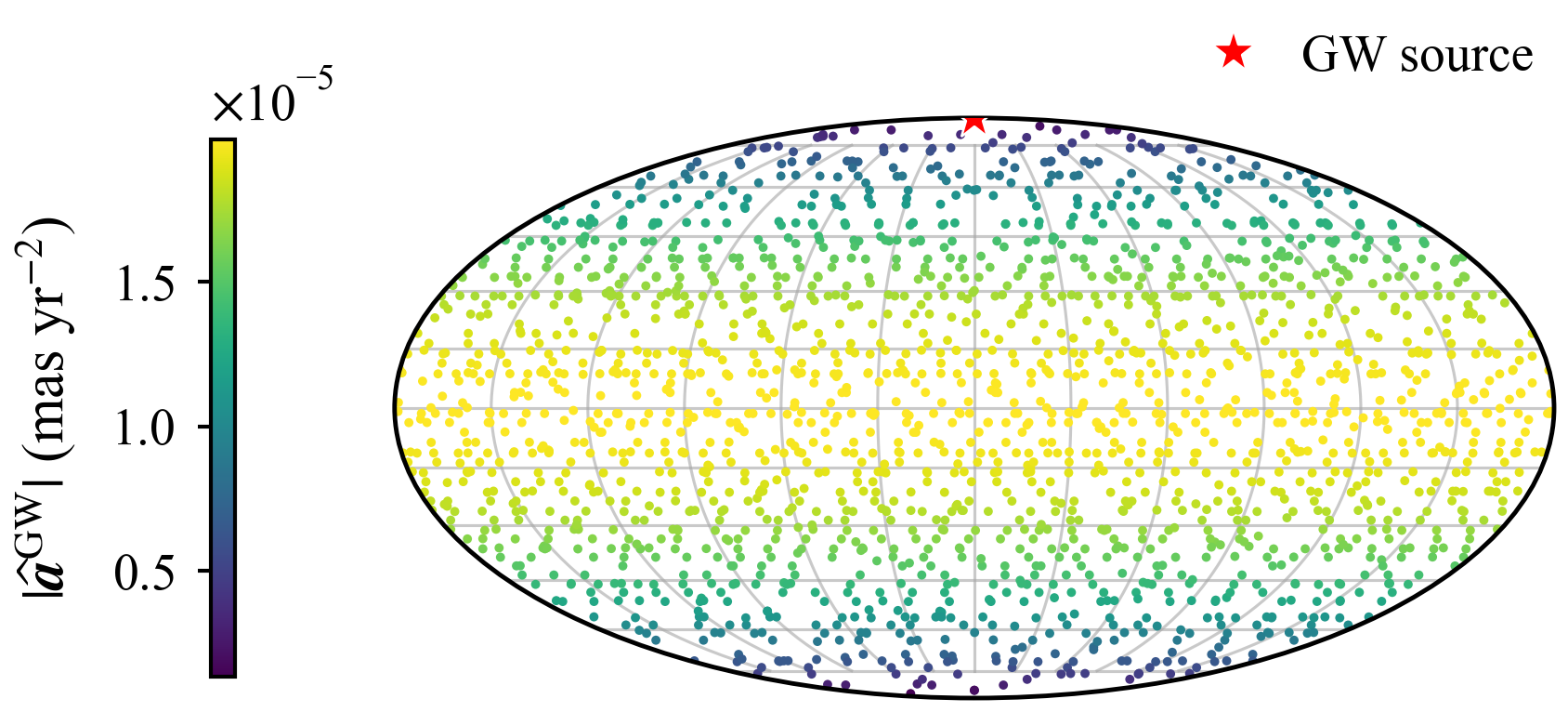}
  \caption{
  Sky distribution of the GW-induced angular acceleration $\widehat{\boldsymbol a}^{\rm GW}$ with $2\times10^3$ stars. The color scale shows its magnitude in $\mathrm{mas\,yr^{-2}}$, and the red star marks the GW source. The monochromatic signal has strain amplitude $A=9.0\times10^{-13}$, frequency $f_{\rm GW}=3.3\times10^{-9}\,\mathrm{Hz}$, orientation of the transverse polarization basis $\psi=0$, inclination $\iota=0$, and initial phase $\phi_0=0$.
  }
  \label{fig:snap}
\end{figure}

Approximating the catalog proper motions by instantaneous derivatives at their reference epochs, the GW-induced proper-motion difference between Gaia DR3 and DR2 is
\begin{equation}
 \Delta\boldsymbol\mu_i^{\rm GW}
 \simeq
 \dot{\boldsymbol\xi}_i^{\rm GW}(t_3)
 -
 \dot{\boldsymbol\xi}_i^{\rm GW}(t_2),
 \label{eq:instantaneous-pm-approx}
\end{equation}
where $t_2$ and $t_3$ are the reference epochs of DR2 and DR3, J2015.5 and J2016.0, respectively. For a GW signal that varies slowly over both observing intervals and their reference-epoch separation, $\Delta T=t_3-t_2=0.5\,{\rm yr}$, the angular-acceleration estimator satisfies
\begin{equation}
 \widehat{\boldsymbol a}_i^{\rm GW}
 \equiv
 \frac{\Delta\boldsymbol\mu_i^{\rm GW}}{\Delta T}
 \simeq
 \boldsymbol a_i^{\rm GW},
 \label{eq:slow-gw-approx}
\end{equation}
where $\boldsymbol a_i^{\rm GW}$ is evaluated at the midpoint of the two reference epochs. This approximation applies to GW periods longer than the observing intervals and the epoch separation; its validity range is discussed below.

\textit{\textbf{Data set and analysis---}}
The matched Gaia DR2-DR3 catalog contains approximately $3\times10^8$ sources. We retain sources with valid astrometry in both releases and apply cuts based on astrometric solution quality, photometric variability, and reported non-single-star solutions, as detailed in the Appendix. We then compress the two-component proper-motion differences of the retained sources into sky-photometry cells. Sources are grouped by HEALPix level 4 \cite{HEALPix}. For each group $g$, we compute
\begin{equation}
P_g=\sum_{i\in g}(C_i^{\rm cal})^{-1},\qquad
 \bar{\boldsymbol d}_g=P_g^{-1}\sum_{i\in g}(C_i^{\rm cal})^{-1}\boldsymbol d_i
 \label{eq:compression}
\end{equation}
where $\boldsymbol d_i$ is the observed two-component difference for source $i$ and $C_i^{\rm cal}$ is its calibrated covariance, defined in the Appendix. The compressed covariance is $C_g=P_g^{-1}$. Every retained source contributes to its group's weighted mean. This approximation requires the model to vary negligibly within each group.

The model for the residual proper motion field is
\begin{equation}
\bar{\boldsymbol d}_i =
\boldsymbol\mu_i^{\rm GW}
+\boldsymbol\mu_i^{\rm rot}
+\boldsymbol\mu_i^{\rm glide}
+\boldsymbol\epsilon_i ,
  \label{eq:model}
\end{equation}
The global frame rotation induces the apparent proper-motion field $\boldsymbol{\mu}^{\rm rot}_i=\boldsymbol{\omega}\times\hat{\boldsymbol n}_i$, which we project onto the local $(\hat{\boldsymbol e}_{\alpha},\hat{\boldsymbol e}_{\delta})$ basis. Here, $\boldsymbol{\omega}$ is the angular-velocity vector describing the rigid rotation of the astrometric reference frame. This field is the $l=1$ toroidal (magnetic-type) vector-spherical-harmonic mode, corresponding to an apparent circulation on the sky. $\boldsymbol{\mu}^{\rm glide}_i=\boldsymbol{g}-(\boldsymbol{g}\cdot\hat{\boldsymbol n}_i)\hat{\boldsymbol n}_i$ is the $l=1$ electric (spheroidal) glide field, where $\boldsymbol{g}$ is the amplitude vector of the dipolar proper-motion pattern: its direction sets the preferred sky direction, and its magnitude sets the pattern’s strength. This expression projects $\boldsymbol{g}$ onto the tangent plane at source $i$.

The parameters ${\boldsymbol \omega}$, ${\boldsymbol \omega}_{b}$, and ${\boldsymbol g}$ are treated as nuisance parameters and marginalized over in the Bayesian analysis. $\boldsymbol\epsilon_i$ denotes the residual proper motion, including measurement noise and unmodeled source-specific effects, and its covariance is described by $C_i^{\rm cal}$. We do not fit a separate constant intrinsic proper motion for each source. Our observable is the DR3--DR2 proper-motion difference divided by the separation of the catalog reference epochs, so a constant intrinsic proper motion cancels to first order. 

Feng et al.~\cite{Feng_2024} reported magnitude- and color-dependent differential astrometric calibration between Gaia DR2 and DR3, including variations in the frame-rotation parameters. Motivated by these results, we let the frame rotation depend on source photometry by jointly binning the sample by $G$ magnitude and BP--RP color. We use three $G$-magnitude intervals, $[0,10.5)$, $[10.5,13)$, and $[13,99)$, and three BP--RP color intervals, $[-5,0.5)$, $[0.5,1.5)$, and $[1.5,5)$. These define at most nine joint photometric bins, and only the bins with enough sources are retained. Each retained bin $b$ is assigned an independent three-component rotation vector $\boldsymbol\omega_b$, giving the field $\boldsymbol{\mu}_i^{\rm rot} =\boldsymbol\omega_{b(i)}\times\hat{\boldsymbol n}_i$. Thus, the binned model contains $3N_{\rm bin}$ rotation nuisance parameters, where $N_{\rm bin}$ is the number of retained bins. For the model that includes both binned rotation and glide, a single glide vector $\boldsymbol g$ is shared across all bins. Comparisons with the global-rotation model assess how magnitude- and BP--RP-color-dependent frame rotation affects the GW search results

\textit{\textbf{Sensitivity Curve and upper limits---}}
We assess the evidence for a GW signal by comparing two hypotheses: $\mathcal{M}_0$, comprising frame rotation and glide, and $\mathcal{M}_1$, which additionally includes a GW signal. For each model, the Bayesian evidence is
\begin{equation}
Z_k \equiv p(\boldsymbol{d}\mid\mathcal{M}_k)
= \int
\mathcal{L}(\boldsymbol{d}\mid\boldsymbol{\theta}_k,\mathcal{M}_k)
\pi(\boldsymbol{\theta}_k\mid\mathcal{M}_k)
\,d\boldsymbol{\theta}_k ,
\end{equation}
where $\boldsymbol{d}$ denotes the data, $\boldsymbol{\theta}_k$ comprises all free parameters, and $\mathcal{L}$ and $\pi$ are the likelihood and prior, respectively. We evaluate the evidence independently using nested sampling as implemented in \texttt{dynesty} \cite{Speagle2020} and quantify the relative support for the GW hypothesis through the Bayes factor,
\begin{equation}
B_{10} \equiv \frac{Z_1}{Z_0}.
\end{equation}
Both models use the same data, covariance model, and priors on the frame-rotation and glide parameters. The GW model additionally includes
$\boldsymbol{\theta}_{\rm GW}=(A,f,\iota, \psi,\phi_0,\alpha_{\rm GW},\delta_{\rm GW})$.

We fit the two-component Gaia DR3--DR2 proper-motion differences with the two models. We use \texttt{dynesty} to compute the Bayesian evidence and sample the posterior. We adopt an isotropic prior for the GW propagation direction, uniform in $\alpha_{\rm GW}$ and $\sin\delta_{\rm GW}$, and an isotropic binary-orientation prior, uniform in $\cos\iota$. The polarization angle $\psi$ and initial phase $\phi_0$ have uniform priors. We adopt priors uniform in $\log_{10}A$ and $\log_{10}f$ for the strain amplitude and GW frequency, respectively. (We also adopt uniform priors for the uncertainty inflation factor $f_G$ and the systematic floor $s_\mu$, both of which are described in the Appendix.)

After calculation, we find $B_{10}\simeq 1$ throughout the analysis, indicating that the data do not significantly favor either hypothesis and provide no compelling evidence for a continuous GW signal. We therefore derive upper limits on the GW strain amplitude from its marginalized posterior under $\mathcal{M}_1$.

Marginalizing over the source inclination, initial phase, sky position, polarization angle, and nuisance parameters yields $p(A\mid\boldsymbol d,f,\mathcal M_i)$. We define the 95\% credible upper limit, $A_{95}$, by
\begin{equation}
\int_{A_{\min}}^{A_{95}(f)}
p(A\mid\boldsymbol d,f,\mathcal M_i)\,{\rm d}A
=0.95,
\label{eq:strain-upper-limit}
\end{equation}
where $A_{\min}=10^{-15}$ is the lower bound of the amplitude prior. The model $\mathcal{M}_i$ denotes one of the following treatments: GW only, GW plus glide, GW plus global rotation, GW plus binned rotation, or GW plus binned rotation and a glide. The resulting $A_{95}(f_{\rm GW})$ therefore quantifies the frequency-dependent upper limit under each treatment of the astrometric systematic effects.

\begin{figure}[t]
  \centering
  \includegraphics[width=\columnwidth]{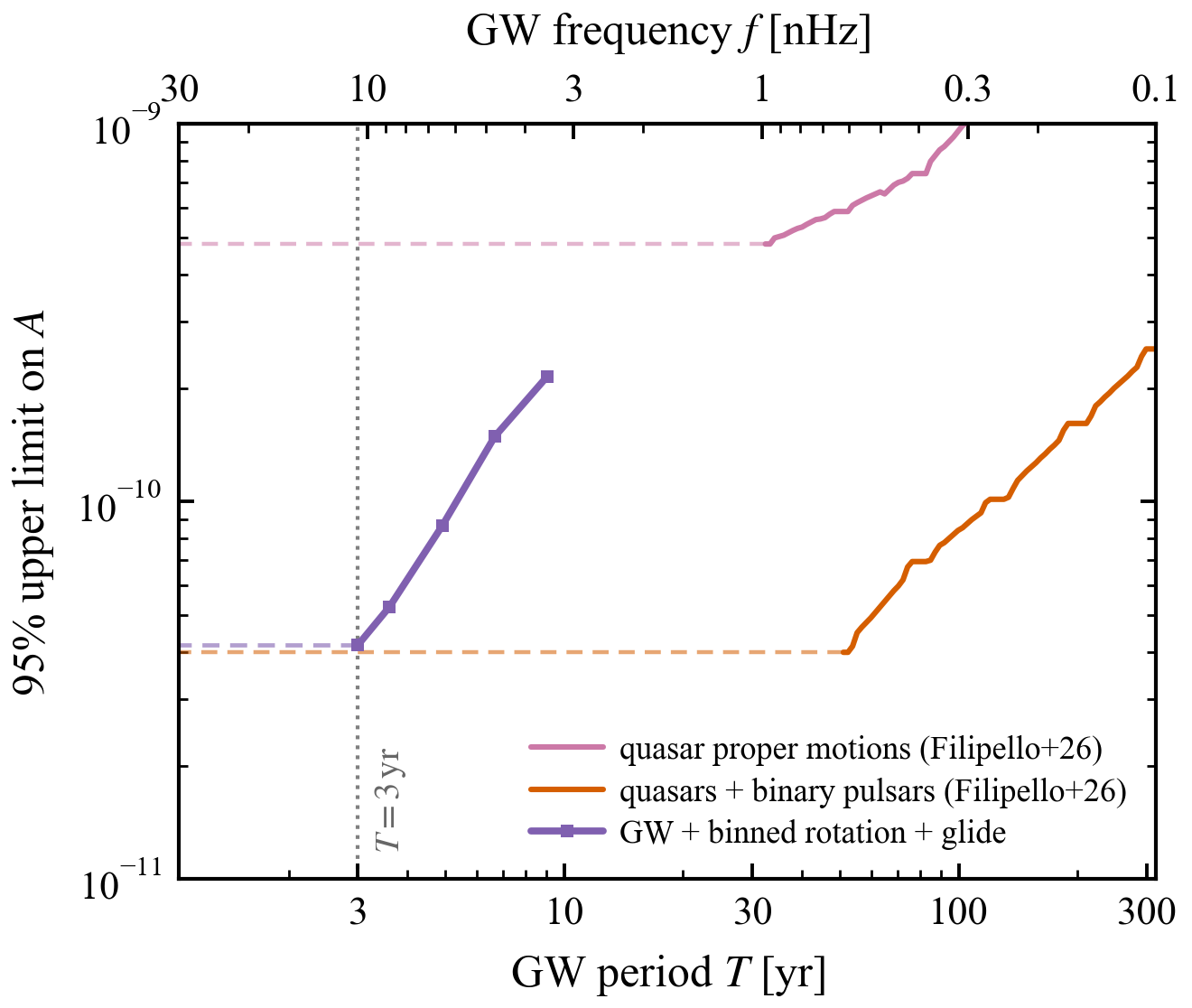}
  \caption{
  Strain upper limits from the Gaia DR3--DR2 acceleration analysis. Purple squares show the 95\% credible limits for the model including magnitude- and color-binned frame rotation and global glide. Pink and orange curves show the constraints from Gaia DR3 quasar proper motions alone and from their joint analysis with binary-pulsar orbital period by Filipello \cite{Filipello_26}. The vertical dotted line marks $T_{\rm obs}=3\,\mathrm{yr}$, the shortest period evaluated in our analysis. Filipello restricts their analysis to $f \sim (4T_{\rm obs})^{-1}$ for their slowly varying signal approximation. Dashed segments continue each curve at its last evaluated value solely for visual comparison and do not represent independently derived limits.
  }
  \label{fig:upper_limit}
\end{figure}

Fig.~\ref{fig:upper_limit} compares our strain upper limits for the binned rotation plus glide model with the Gaia DR3 quasar-only and joint astrometry-pulsar constraints of Filipello et al.~\cite{Filipello_26}. We evaluate our limits for GW periods of 3-10 yr. The three-year boundary is comparable to the characteristic timescale of the Gaia DR3 observations. Moore et al.~\cite{Moore2017} similarly identify $f\sim1/T$ as a characteristic frequency for astrometric GW searches, although their result focuses on the proper motion rather than the proper motion difference observable used here. At periods longer than 10 yr, the acceleration response weakens rapidly, reducing the sensitivity of the present data; as we can see, our result will be worse than the Quasar result (pink). Therefore, we focus on the 3-10 yr period for the acceleration method.

For periods shorter than $T_{\rm obs}\sim 3\,\mathrm{yr}$, signal variation within the observing windows may attenuate its contribution to the fitted proper motions, which means our method may be unsuitable anymore. Filipello adopted a maximum frequency $f_{\max}=(4T_{\rm obs})^{-1}$ for their slowly varying signal approximation, so their published curves end at a lower frequency than ours. To aid visual comparison, we extend their curves toward shorter periods and ours toward periods below $3\,\mathrm{yr}$, using dashed horizontal segments. These extensions hold each curve at its last evaluated value and do not represent independently derived limits. Near $P_{\rm GW}=3\,\mathrm{yr}$, our limit lies below their quasar-only limit and approaches their joint-analysis limit.

We also find that the strain upper limits obtained with different methods exhibit distinct log–log slopes as a function of GW period in Fig.~\ref{fig:upper_limit}. The quasar proper-motion and joint-analysis curves have log-log slopes close to $+1$, consistent with a proper-motion response proportional to $h_0 f$. In the regime where the acceleration approximation is valid, the angular-acceleration response scales as $h_0 f^2$. The acceleration-based limits (purple) thus decrease more rapidly toward shorter periods and are better than the Quasar limits (pink) near $P_{\rm GW}\simeq3\mathrm{yr}$.

\begin{figure}[t]
  \centering
  \includegraphics[width=\columnwidth]{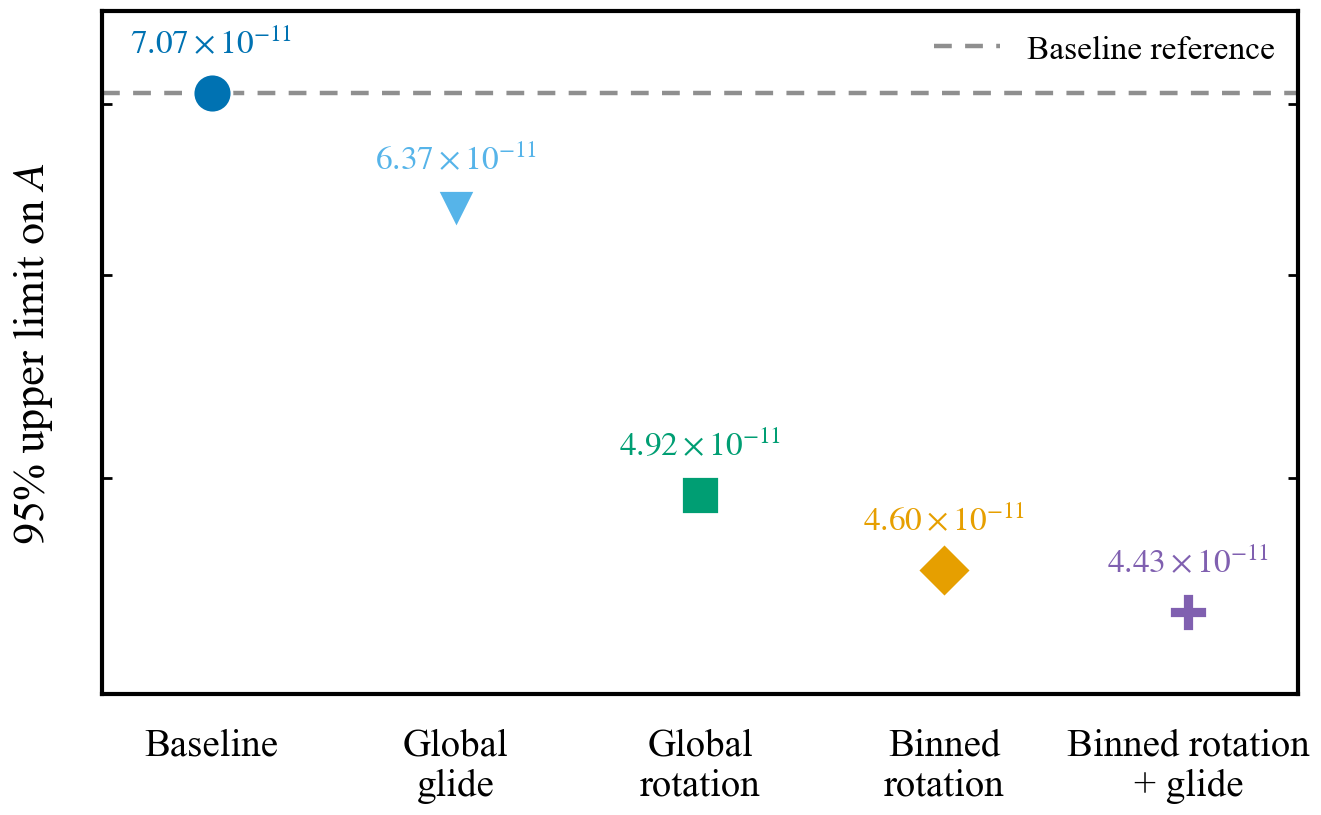}
    \caption{Conditional 95\% upper limits $A_{95}$ on the GW strain amplitude $A$ at fixed frequency $f_{\rm GW}=\text{10nHz}$ for different treatments of astrometric systematics. The markers show the limits obtained with the baseline, global-glide, global-rotation, binned-rotation, and binned-rotation-plus-glide models; the gray dashed line indicates the baseline result. The numerical values are shown above the markers.
    }
  \label{fig:compare_effets}
\end{figure}

Figure~\ref{fig:compare_effets} compares five models for astrometric systematics, each fitted jointly with the GW-induced astrometric response: a baseline model with no systematic terms, a global glide, a global frame rotation, a magnitude- and color-binned frame rotation, and a binned frame rotation combined with a global glide. The dots represent the 95\% upper limit of strain for different models at $f=10\mathrm{nHz}$.

Frame rotation is the toroidal ($l=1$ B-mode) component of the astrometric vector field and can arise from differential spins of the DR2 and DR3 reference frames. Glide is the spheroidal ($l=1$ E-mode) component. In a single-catalog proper-motion field, glide can be associated with secular aberration drift, caused by the acceleration of the Solar System barycenter in the Galactic gravitational potential, or with distance-dependent secular parallax, caused by the barycenter's motion relative to the source population. Because our method is the DR3--DR2 proper motion difference, however, a fitted glide term instead represents a residual dipolar difference between the two releases and need not correspond directly to either physical effect. Under uniform full-sky weighting, these dipolar patterns are orthogonal to the leading GW multipoles. When sky coverage is nonuniform and covariance weighting is applied in the Gaia sample, this orthogonality can break down, allowing the fitted systematics to couple to the GW response and alter the inferred strain limits. Figure~\ref{fig:compare_effets} shows that the rotation model yields a tighter upper limit than the glide model, suggesting that rotation-like residuals have a greater effect on the GW-amplitude posterior in this sample. Allowing the frame rotation to vary with the $G$ magnitude and BP--RP color further tightens the limit, consistent with source-dependent differential frame calibration \cite{Feng_2024}.

\textit{\textbf{Conclusion---}}
In this Letter, we have presented the first search for individual continuous gravitational waves using angular-acceleration estimates derived from proper-motion differences between Gaia DR2 and DR3 stars. The additional information from two catalog releases complements searches based on a single release's proper motion field. For GW periods comparable to or longer than the observing baseline, $f_{\rm GW}\lesssim T_{\rm obs}^{-1}$, the GW response scales as $h_0 f^{2}$, compared with $h_0 f$ for a proper-motion method. This gives the acceleration-based method a more rapidly improving strain response toward shorter periods. We find no significant evidence for a GW signal and derive 95\% credible upper limits, reaching an order of $4\times10^{-11}$ near $10\,\mathrm{nHz}$. 

The upper limits depend on how the astrometric systematics are modeled. We jointly fit reference-frame rotation and glide, allowing the rotation to vary with $G$ magnitude and ${\rm BP}-{\rm RP}$ color. Although these dipolar fields are orthogonal to the leading GW multipoles for uniformly weighted full-sky data, the nonuniform sky coverage and covariance weights of the Gaia sample can couple their templates to the GW response. The model with binned-rotation and glide yields the tightest observed limit among the models considered. This model dependence shows that differential frame systematics must be assessed when interpreting astrometric GW constraints from successive Gaia releases.

Applying our 95\% strain upper limit to OJ~287 using the relation in Ref.~\cite{Chen2026}, we infer a corresponding upper limit on its intrinsic chirp mass. This yields $\mathcal{M}_{c}\approx1.2\times10^{12}\,M_\odot$, roughly three orders of magnitude above the reference estimate for OJ~287 \cite{Dey2018OJ287, Valtonen2016OJ287}. This is because our acceleration method cannot be applied to high frequencies, which limits the results. To extend the analysis to higher frequencies, where the acceleration method breaks down, we are developing a model that incorporates the Gaia scanning law and predicts the GW-induced perturbations to the reported positions, parallaxes, and proper motions in each release. This approach is expected to enable reliable GW constraints at higher frequencies.

Gaia DR4 will offer an opportunity to strengthen these searches through its longer observing baseline and improved astrometric precision. The model dependence illustrated in Fig.~\ref{fig:compare_effets} demonstrates that comparisons among alternative prescriptions for frame rotation and glide will remain necessary in the analysis of DR4 data. In particular, we must jointly assess magnitude- and color-dependent rotation-calibration effects with the GW signal to establish reliable strain constraints and evaluate potential signal candidates.

\textit{\textbf{Acknowledgments---}} 
This work is supported by the National Key R\&D Program of China (grant No. 2024YFC2207700). H.T. is supported by the National Science and Technology Major Project of China (No. 2024ZD1100601). This work made use of the High Performance Computing Resource in the Core Facility for Advanced Research Computing at Shanghai Astronomical Observatory.

\section{Appendix}
\textit{\textbf{Data selection criterion---}}
We require finite positions, parallaxes, and both proper-motion components in each release, together with finite and strictly positive formal uncertainties for these quantities. These cuts ensure that the DR3--DR2 proper-motion difference and its formal covariance can be constructed.

We assess astrometric quality using the solution type and the renormalized unit weight error (RUWE). A five-parameter solution fits two position components, parallax, and two proper-motion components; a six-parameter solution also estimates a pseudocolor. We retain sources identified by the quality flag in our input catalog as having a five-parameter solution. RUWE measures the goodness of fit of the single-star model to the along-scan observations. Values near unity indicate a fit consistent with the expected scatter, whereas larger values indicate a poorer fit. We require $\mathrm{RUWE}<1.1$ for the RUWE field provided in the input catalog.

To reduce contamination from potentially time-dependent source motion, we exclude sources with a reported non-single-star solution. We also exclude sources flagged as photometric variables or assigned a variability class in the DR3 fields used for selection. An absent solution or classification is treated as the absence of a reported flag, not as confirmation that a source is single or nonvariable.

\textit{\textbf{Covariance calibration for DR2--DR3---}}
For release $r$, the catalog-level covariance of the two-component proper motion is

\begin{equation}
 C_{r,i}=
 \begin{pmatrix}
 \sigma_{\alpha*,r,i}^{2} &
 \rho_{\alpha\delta,r,i}\sigma_{\alpha*,r,i}\sigma_{\delta,r,i}\\
 \rho_{\alpha\delta,r,i}\sigma_{\alpha*,r,i}\sigma_{\delta,r,i} &
 \sigma_{\delta,r,i}^{2}
 \end{pmatrix}.
 \label{eq:catalog-cov}
\end{equation}

Writing $C_{r,i}$ for release $r\in\{2,3\}$ and $C_{23,i}$ for their cross-release covariance, the difference covariance is
\begin{equation}
 C_{\Delta,i}=C_{3,i}+C_{2,i}-C_{23,i}-C_{23,i}^{T}.
 \label{eq:delta-cov}
\end{equation}

Because Gaia DR2 and DR3 share observations and calibration parameters, their proper-motion errors are correlated. The cross-release covariance $C_{23,i}$ cannot be determined from the within-release correlation coefficients alone. In the absence of an external estimate, we neglect this term and adopt $C_{\Delta,i}\simeq C_{3,i}+C_{2,i}$ throughout our analysis.

We calibrate the formal proper-motion covariance using a magnitude-dependent uncertainty inflation factor $f_G(G_i)$ and an additive systematic floor $s_\mu$:
\begin{equation}
 C_i^{\rm cal}
 =
 f_G(G_i)^2 C_{\Delta,i}+s_\mu^2 I_2,
 \label{eq:error-calibration}
\end{equation}
where $I_2$ is the $2\times2$ identity matrix. The factor $f_G$ rescales the formal standard deviations, while $s_\mu$ adds an equal variance to the two proper-motion components without introducing an off-diagonal covariance.

Previous studies have shown that Gaia proper-motion uncertainties can require sample-dependent inflation \cite{brandt2021hgca}, while spatially correlated systematics can limit the precision of mean proper motions \cite{vasiliev2021clusters}. We therefore infer the calibration parameters $f_G$ and $s_\mu$ from our sample, jointly with the GW and astrometric nuisance parameters, rather than adopting values calibrated for other source populations.

\bibliographystyle{apsrev4-2}
\bibliography{references}

@misc{LISA2024,
  author        = {Colpi, Monica and Danzmann, Karsten and
                   Hewitson, Martin and others},
  title         = {{LISA Definition Study Report}},
  year          = {2024},
  eprint        = {2402.07571},
  archivePrefix = {arXiv},
  primaryClass  = {astro-ph.CO},
  doi           = {10.48550/arXiv.2402.07571}
}

@article{NANOGrav2023,
  author        = {Agazie, Gabriella and Anumarlapudi, Akash and
                   Archibald, Anne M. and others},
  collaboration = {NANOGrav},
  title         = {The {NANOGrav} 15-year Data Set:
                   Evidence for a Gravitational-Wave Background},
  journal       = {Astrophys. J. Lett.},
  volume        = {951},
  number        = {1},
  pages         = {L8},
  year          = {2023},
  doi           = {10.3847/2041-8213/acdac6},
  eprint        = {2306.16213},
  archivePrefix = {arXiv},
  primaryClass  = {astro-ph.HE}
}

@article{BookFlanagan2011,
  author        = {Book, Laura G. and Flanagan, Eanna E.},
  title         = {Astrometric Effects of a Stochastic
                   Gravitational Wave Background},
  journal       = {Phys. Rev. D},
  volume        = {83},
  pages         = {024024},
  year          = {2011},
  doi           = {10.1103/PhysRevD.83.024024}
}

@article{Moore2017,
  author        = {Moore, Christopher J. and Mihaylov, Deyan P. and
                   Lasenby, Anthony and Gilmore, Gerard},
  title         = {Astrometric Search Method for Individually
                   Resolvable Gravitational Wave Sources with {Gaia}},
  journal       = {Phys. Rev. Lett.},
  volume        = {119},
  pages         = {261102},
  year          = {2017},
  doi           = {10.1103/PhysRevLett.119.261102}
}

@article{Klioner2018,
  author        = {Klioner, Sergei A.},
  title         = {{Gaia}-like Astrometry and Gravitational Waves},
  journal       = {Class. Quantum Grav.},
  volume        = {35},
  number        = {4},
  pages         = {045005},
  year          = {2018},
  doi           = {10.1088/1361-6382/aa9f57},
  eprint        = {1710.11474},
  archivePrefix = {arXiv},
  primaryClass  = {astro-ph.HE}
}

@article{Jaraba2023,
  author        = {Jaraba, Santiago and Garc{\'i}a-Bellido, Juan and
                   Kuroyanagi, Sachiko and Ferraiuolo, Sarah and
                   Braglia, Matteo},
  title         = {Stochastic Gravitational Wave Background
                   Constraints from {Gaia DR3} Astrometry},
  journal       = {Mon. Not. R. Astron. Soc.},
  volume        = {524},
  number        = {3},
  pages         = {3609--3622},
  year          = {2023},
  doi           = {10.1093/mnras/stad2141},
  eprint        = {2304.06350},
  archivePrefix = {arXiv},
  primaryClass  = {astro-ph.CO}
}

@article{Darling2025,
  author        = {Darling, Jeremy},
  title         = {A New Approach to the Low-frequency Stochastic
                   Gravitational-wave Background: Constraints from
                   Quasars and the Astrometric {Hellings--Downs} Curve},
  journal       = {Astrophys. J. Lett.},
  volume        = {982},
  number        = {2},
  pages         = {L46},
  year          = {2025},
  doi           = {10.3847/2041-8213/adbf0d},
  eprint        = {2412.08605},
  archivePrefix = {arXiv},
  primaryClass  = {astro-ph.CO}
}

@article{GW150914,
  title = {Observation of Gravitational Waves from a Binary Black Hole Merger},
  author = {Abbott, B. P. and others},
  collaboration = {LIGO Scientific Collaboration and Virgo Collaboration},
  journal = {Phys. Rev. Lett.},
  volume = {116},
  issue = {6},
  pages = {061102},
  numpages = {16},
  year = {2016},
  month = {Feb},
  publisher = {American Physical Society},
  doi = {10.1103/PhysRevLett.116.061102},
  url = {https://link.aps.org/doi/10.1103/PhysRevLett.116.061102}
}

@article{Feng_2024,
doi = {10.3847/1538-4365/ad27d2},
url = {https://doi.org/10.3847/1538-4365/ad27d2},
year = {2024},
month = {mar},
publisher = {The American Astronomical Society},
volume = {271},
number = {2},
pages = {50},
author = {Feng, Fabo and Rui, Yicheng and Xuan, Yifan and Jones, Hugh},
title = {Modeling and Calibration of Gaia, Hipparcos, and Tycho-2 Astrometric Data for the Detection of Dark Companions},
journal = {The Astrophysical Journal Supplement Series}
}

@misc{Filipello_26,
      title={Combining astrometry with pulsar timing: the first joint analysis of very low frequency gravitational waves}, 
      author={Lorenzo Filipello and William Beordo and Volodymyr Akhmetov and Mariateresa Crosta and Mario Gilberto Lattanzi},
      year={2026},
      eprint={2609.21606},
      archivePrefix={arXiv},
      primaryClass={astro-ph.IM},
      url={https://arxiv.org/abs/2609.21606}, 
}

@article{Jaffe2003,
  author  = {Jaffe, Andrew H. and Backer, Donald C.},
  title   = {Gravitational Waves Probe the Coalescence Rate of Massive Black Hole Binaries},
  journal = {The Astrophysical Journal},
  volume  = {583},
  pages   = {616--631},
  year    = {2003},
  doi     = {10.1086/345443}
}

@article{Wyithe2003,
  author  = {Wyithe, J. Stuart B. and Loeb, Abraham},
  title   = {Low-Frequency Gravitational Waves from Massive Black Hole Binaries: Predictions for {LISA} and Pulsar Timing Arrays},
  journal = {The Astrophysical Journal},
  volume  = {590},
  pages   = {691--706},
  year    = {2003},
  doi     = {10.1086/375187}
}

@article{Enoki2004,
  author  = {Enoki, Motohiro and Inoue, Kaiki T. and Nagashima, Masahiro and Sugiyama, Naoshi},
  title   = {Gravitational Waves from Supermassive Black Hole Coalescence in a Hierarchical Galaxy Formation Model},
  journal = {The Astrophysical Journal},
  volume  = {615},
  pages   = {19--28},
  year    = {2004},
  doi     = {10.1086/424475}
}

@article{Sesana2008,
  author  = {Sesana, Alberto and Vecchio, Alberto and Colacino, Carlo Nicola},
  title   = {The stochastic gravitational-wave background from massive black hole binary systems: implications for observations with Pulsar Timing Arrays},
  journal = {Monthly Notices of the Royal Astronomical Society},
  volume  = {390},
  pages   = {192--209},
  year    = {2008},
  doi     = {10.1111/j.1365-2966.2008.13682.x}
}

@article{Zic2023PPTADR3,
  author  = {Zic, Andrew and others},
  title   = {The {Parkes Pulsar Timing Array} third data release},
  journal = {Publications of the Astronomical Society of Australia},
  volume  = {40},
  pages   = {e049},
  year    = {2023},
  doi     = {10.1017/pasa.2023.36}
}

@article{Agazie2023Timing,
  author  = {Agazie, Gabriella and others},
  title   = {The {NANOGrav} 15 yr Data Set: Observations and Timing of 68 Millisecond Pulsars},
  journal = {The Astrophysical Journal Letters},
  volume  = {951},
  pages   = {L9},
  year    = {2023},
  doi     = {10.3847/2041-8213/acda9a}
}

@article{Antoniadis2023EPTAData,
  author  = {Antoniadis, J. and others},
  title   = {The second data release from the {European Pulsar Timing Array}. {I}. The dataset and timing analysis},
  journal = {Astronomy \& Astrophysics},
  volume  = {678},
  pages   = {A48},
  year    = {2023},
  doi     = {10.1051/0004-6361/202346841}
}

@article{Rana2025InPTADR2,
  author  = {Rana, Prerna and others},
  title   = {The {Indian Pulsar Timing Array} data release 2: {I}. Dataset and timing analysis},
  journal = {Publications of the Astronomical Society of Australia},
  volume  = {42},
  pages   = {e108},
  year    = {2025},
  doi     = {10.1017/pasa.2025.10066}
}

@article{Chen2025CPTANoise,
  author  = {Chen, Siyuan and others},
  title   = {The {Chinese Pulsar Timing Array} data release {I}. Single pulsar noise analysis},
  journal = {Astronomy \& Astrophysics},
  volume  = {699},
  pages   = {A165},
  year    = {2025},
  doi     = {10.1051/0004-6361/202452550}
}

@article{Miles2025MPTAData,
  author  = {Miles, Matthew T. and others},
  title   = {The {MeerKAT Pulsar Timing Array}: the 4.5-yr data release and the noise and stochastic signals of the millisecond pulsar population},
  journal = {Monthly Notices of the Royal Astronomical Society},
  volume  = {536},
  pages   = {1467--1488},
  year    = {2025},
  doi     = {10.1093/mnras/stae2572}
}

@article{Agazie2023GWB,
  author  = {Agazie, Gabriella and others},
  title   = {The {NANOGrav} 15 yr Data Set: Evidence for a Gravitational-wave Background},
  journal = {The Astrophysical Journal Letters},
  volume  = {951},
  pages   = {L8},
  year    = {2023},
  doi     = {10.3847/2041-8213/acdac6}
}

@article{Reardon2023GWB,
  author  = {Reardon, Daniel J. and others},
  title   = {Search for an Isotropic Gravitational-wave Background with the {Parkes Pulsar Timing Array}},
  journal = {The Astrophysical Journal Letters},
  volume  = {951},
  pages   = {L6},
  year    = {2023},
  doi     = {10.3847/2041-8213/acdd02}
}

@article{Antoniadis2023EPTAGWB,
  author  = {Antoniadis, J. and others},
  title   = {The second data release from the {European Pulsar Timing Array}. {III}. Search for gravitational wave signals},
  journal = {Astronomy \& Astrophysics},
  volume  = {678},
  pages   = {A50},
  year    = {2023},
  doi     = {10.1051/0004-6361/202346844}
}

@article{Xu2023CPTAGWB,
  author  = {Xu, Heng and others},
  title   = {Searching for the Nano-Hertz Stochastic Gravitational Wave Background with the {Chinese Pulsar Timing Array} Data Release {I}},
  journal = {Research in Astronomy and Astrophysics},
  volume  = {23},
  pages   = {075024},
  year    = {2023},
  doi     = {10.1088/1674-4527/acdfa5}
}

@article{Miles2025MPTAGWB,
  author  = {Miles, Matthew T. and others},
  title   = {The {MeerKAT Pulsar Timing Array}: the first search for gravitational waves with the {MeerKAT} radio telescope},
  journal = {Monthly Notices of the Royal Astronomical Society},
  volume  = {536},
  pages   = {1489--1500},
  year    = {2025},
  doi     = {10.1093/mnras/stae2571}
}

@article{Chen2026,
  title = {Nanohertz Gravitational-Wave Constraints on Supermassive Binary Black Holes at Cosmic Dawn},
  author = {Chen, Yiqin and Zhao, Shi-Yi and Zhu, Xingjiang and Bhat, N. D. Ramesh and Cardinal Tremblay, Jacob and Cury{\l{}}o, Ma{\l{}}gorzata and Di Marco, Valentina and Hobbs, George and Ling, Wenhua and Mandow, Rami F. and Manchester, Richard N. and Mishra, Saurav and Reardon, Daniel J. and Roch, Sparrow and Russell, Christopher J. and Shannon, Ryan M. and Wang, Jingbo and Wang, Shuangqiang and Zic, Andrew},
  collaboration = {PPTA Collaboration},
  journal = {Phys. Rev. Lett.},
  volume = {137},
  issue = {19},
  pages = {191401},
  numpages = {7},
  year = {2026},
  month = {Nov},
  publisher = {American Physical Society},
  doi = {10.1103/5z5g-shly},
  url = {https://link.aps.org/doi/10.1103/5z5g-shly}
}

@article{Babak2016CW,
  author  = {Babak, Stanislav and others},
  title   = {{European Pulsar Timing Array} limits on continuous gravitational waves from individual supermassive black hole binaries},
  journal = {Monthly Notices of the Royal Astronomical Society},
  volume  = {455},
  number  = {2},
  pages   = {1665--1679},
  year    = {2016},
  doi     = {10.1093/mnras/stv2092},
  eprint  = {1509.02165},
  archivePrefix = {arXiv},
  primaryClass  = {astro-ph.CO}
}

@article{Agazie2023CW,
  author  = {Agazie, Gabriella and others},
  title   = {The {NANOGrav} 15 yr Data Set: Bayesian Limits on Gravitational Waves from Individual Supermassive Black Hole Binaries},
  journal = {The Astrophysical Journal Letters},
  volume  = {951},
  number  = {2},
  pages   = {L50},
  year    = {2023},
  doi     = {10.3847/2041-8213/ace18a},
  eprint  = {2306.16222},
  archivePrefix = {arXiv},
  primaryClass  = {astro-ph.HE}
}

@article{Falxa2023CW,
  author  = {Falxa, M. and others},
  title   = {Searching for continuous gravitational waves in the second data release of the {International Pulsar Timing Array}},
  journal = {Monthly Notices of the Royal Astronomical Society},
  volume  = {521},
  number  = {4},
  pages   = {5077--5086},
  year    = {2023},
  doi     = {10.1093/mnras/stad812},
  eprint  = {2303.10767},
  archivePrefix = {arXiv},
  primaryClass  = {gr-qc}
}

@article{Zhao2025CW,
  author  = {Zhao, Shi-Yi and others},
  title   = {Searching for Continuous Gravitational Waves in the {Parkes Pulsar Timing Array} Data Release 3},
  journal = {The Astrophysical Journal},
  volume  = {992},
  pages   = {181},
  year    = {2025},
  doi     = {10.3847/1538-4357/ae0719},
  eprint  = {2508.13944},
  archivePrefix = {arXiv},
  primaryClass  = {gr-qc}
}

@article{HEALPix,
  author  = {G{\'o}rski, K. M. and Hivon, E. and Banday, A. J.
             and Wandelt, B. D. and Hansen, F. K.
             and Reinecke, M. and Bartelmann, M.},
  title   = {{HEALPix}: A Framework for High-Resolution Discretization
             and Fast Analysis of Data Distributed on the Sphere},
  journal = {The Astrophysical Journal},
  volume  = {622},
  pages   = {759--771},
  year    = {2005},
  doi     = {10.1086/427976},
  eprint  = {astro-ph/0409513},
  archivePrefix = {arXiv}
}

@article{Braginsky1990,
  author = {Braginsky, V. B. and Kardashev, N. S. and Polnarev, A. G. and Novikov, I. D.},
  journal = {Nuovo Cimento B},
  volume = {105},
  pages = {1141},
  year = {1990}
}

@article{Pyne1996,
  author = {Pyne, T. and Gwinn, C. R. and Birkinshaw, M. and Eubanks, T. Marshall and Matsakis, D. N.},
  journal = {Astrophys. J.},
  volume = {465},
  pages = {566},
  year = {1996}
}

@article{brandt2021hgca,
  author  = {Brandt, Timothy D.},
  title   = {The {Hipparcos}--{Gaia} Catalog of Accelerations:
             {Gaia} {EDR3} Edition},
  journal = {The Astrophysical Journal Supplement Series},
  year    = {2021},
  volume  = {254},
  pages   = {42},
  doi     = {10.3847/1538-4365/abf93c}
}

@article{vasiliev2021clusters,
  author  = {Vasiliev, Eugene and Baumgardt, Holger},
  title   = {{Gaia} {EDR3} view on galactic globular clusters},
  journal = {Monthly Notices of the Royal Astronomical Society},
  year    = {2021},
  volume  = {505},
  number  = {4},
  pages   = {5978--6002},
  doi     = {10.1093/mnras/stab1475}
}

@misc{LIGO_7,
  author        = {{The LIGO Scientific Collaboration} and
                   {The Virgo Collaboration} and
                   {The KAGRA Collaboration}},
  title         = {{GWTC-5.0}: Observations from the Second Part
                   of the Fourth {LIGO}--{Virgo}--{KAGRA}
                   Observing Run and Updates to the
                   Gravitational-Wave Transient Catalog},
  year          = {2026},
  eprint        = {2605.27225},
  archivePrefix = {arXiv},
  primaryClass  = {gr-qc}
}

@article{Speagle2020,
  author  = {Speagle, Joshua S.},
  title   = {{dynesty}: a dynamic nested sampling package for
             estimating Bayesian posteriors and evidences},
  journal = {Monthly Notices of the Royal Astronomical Society},
  volume  = {493},
  number  = {3},
  pages   = {3132--3158},
  year    = {2020},
  doi     = {10.1093/mnras/staa278},
  eprint  = {1904.02180}
}

@article{Dey2018OJ287,
  author  = {Dey, Lankeswar and Valtonen, M. J. and Gopakumar, A. and
             Zola, S. and Hudec, R. and Pihajoki, P. and Ciprini, S. and
             Matsumoto, K. and Sadakane, K. and Kidger, M. and Nilsson, K. and
             others},
  title   = {Authenticating the Presence of a Relativistic Massive Black Hole
             Binary in {OJ} 287 Using Its General Relativity Centenary Flare:
             Improved Orbital Parameters},
  journal = {The Astrophysical Journal},
  volume  = {866},
  number  = {1},
  pages   = {11},
  year    = {2018},
  doi     = {10.3847/1538-4357/aadd95}
}

@article{Valtonen2016OJ287,
  author  = {Valtonen, M. J. and Zola, S. and Ciprini, S. and Gopakumar, A. and
             Matsumoto, K. and Sadakane, K. and Kidger, M. and Gazeas, K. and
             Nilsson, K. and Berdyugin, A. and others},
  title   = {Primary Black Hole Spin in {OJ} 287 as Determined by the General
             Relativity Centenary Flare},
  journal = {The Astrophysical Journal Letters},
  volume  = {819},
  number  = {2},
  pages   = {L37},
  year    = {2016},
  doi     = {10.3847/2041-8205/819/2/L37}
}

\end{document}